\documentclass[twocolumn,pra,superscriptaddress]{revtex4}
\usepackage{amsmath,amssymb,}
\usepackage{graphicx}
\usepackage{dcolumn}
\usepackage{bm}

\usepackage[colorlinks,citecolor=red]{hyperref}

\usepackage{rotating}
\usepackage{floatrow}

\begin{document}

\title{Floquet engineering and simulating exceptional rings with a quantum spin system}

\author{Peng He}\email{hep@smail.nju.edu.cn}
\author{Ze-Hao Huang}
\affiliation{National Laboratory of Solid State Microstructures
and School of Physics, Nanjing University, Nanjing 210093, China}



\date{\today}

\begin{abstract}
Time-periodic driving in the form of coherent radiation provides a powerful tool for the manipulation of topological materials or synthetic quantum matter. In this paper we propose a scheme to realize non-Hermitian semimetals exhibiting exceptional rings in the spectra through Floquet engineering. A transition from a concentric pair of the rings to a dipolar pair is observed. The concentric pair carries only a quantized Berry phase while the dipolar pair possesses opposite Chern numbers in addition, signaling a topological Lifshitz transition of the Fermi surface. The transport properties of the system are addressed, and we find that this transition process is accompanied by the emergence of a nontrivial Hall conductivity. Furthermore, we explore the quantum simulation of non-Hermitian semimetals with a quantum spin system and the characterization of the topology via the long-time dynamics.
\end{abstract}

\maketitle

\emph{Introduction.--} Exceptional points (EPs) are defective degeneracies in the energy spectra where the eigenstates of the Hamiltonian do not span a complete Hilbert space \cite{Bender1998,BergholtzET2019,Ghatak2019}. They are fascinating instances of non-Hermitian (NH) systems emerging as effective descriptions of nonconserved systems such as solids with finite quasi-particle lifetimes \cite{Kozii2017,Yoshida2018,Kimura2019,Michishita2020}, disordered Dirac fermions \cite{Zyuzin2018,Papaj2019}, and artificial lattices with gain and loss or nonreciprocity \cite{Klaiman2008,Cerjan2019,Zhou2018,JLi2019,HJiang2019,DWZhang2020}. Recently, the existence of different exceptional solutions, such as isolated EPs \cite{Yokomizo2020}, exceptional rings (ERs) \cite{XuDuan2017,Kawabata2019,ZhangHJ2019,Kunst2019}, exceptional links \cite{Bergholtz2018,Hu2019,Bergholtz2019}, and exceptional surfaces \cite{ZhangZhou2019,Yamamoto2019,Okugawa2019} in NH semimetals and their topological nature have been recognized \cite{Leykam2017,ShenFu2018,Gong2019,KawabataSymm2019,Zhang2019}. The intriguing features of the topology include that new symmetries \cite{Gong2019} and topology invariants such as energy vorticity \cite{ShenFu2018} and discriminant number \cite{ZYang2019}, anomalous skin effects \cite{ZWang2018,FSong2019,LiGong2019,CHLee2019,LeeHybrid2019,OkumaTopo2019}, and the breakdown of  the usual bulk-boundary correspondence \cite{ZWang2018,YaoNHChern2018,KunstBio2018,Murakami2019,CFang2019}.
Furthermore, certain discrete symmetries are required to stabilize the 1D line nodes of a  3D Hermitian topological semimetal as a result of Bott periodicity for even-odd dichotomy \cite{Burkov2011,Moroz2018}, while the exceptional line in an NH semimetal is stable even without the imposition of any protecting symmetry \cite{Bergholtz2018,Hu2019}.

For conventional topological materials  or synthetic quantum matter \cite{DWZhang2018}, a change in the structure of Fermi surface may happen when an external field is applied. For instance, the Dirac points in graphene can move under the changes of control parameters \cite{Pereira2009,Tarruell2012,DWZhang2012,SLZhu2007}; periodical drives can give rise to a dimension deformation from nodal line to two Weyl points each attached with a $\mathbb{Z}$-monopole charge \cite{Wang2016,Narayan2016,LiGong2018,Salerno2019}, incorporating the coherent manipulation of topological semimetals into the various contexts of Floquet topological matters \cite{Eckardt2017,SYao2017}. Furthermore, the non-Hermiticity has also been included in the field of Floquet states in topological insulators and superconductors \cite{LZhou2018,LZhou2020,HWu2020}.  In this paper, we propose the realization of exceptional-ring semimetals by generalizing the Floquet-engineering methods developed in Refs. \cite{Wang2016,Narayan2016,LiGong2018,Salerno2019,GLiu2012} to NH cases. The NH semimetal exhibiting double ERs is driven by a circularly polarized light, and a transition of the ERs from a concentric pair of which each one carries only a nontrivial Berry phase to a dipolar pair with opposite Chern numbers is observed. This signals an exotic topological Lifshitz transition which has no Hermitian counterpart. Furthermore, we find that this transition process is accompanied by the criticality of the Hall conductivity, which provides experimentally observable evidence for the change of the Fermi surface.

On the other hand, the experimental demonstration and quantum simulation of the NH quantum mechanics has been attracting considerable interests in recent years. Particularly promising approaches using different platforms have been successfully performed. One line of works is engineering the Lindblad master equation using acoustics \cite{ZhuZhang2014,Popa2014}, optical \cite{Zhou2018,Cerjan2019,Xiao2019,KWang2019}, or atomic \cite{SDiehl2011,JLi2019,DWZhang2018,Takasu2020} systems. Other novel protocols adopt alternative methods with no need of controlling an open system \cite{Samsonov2008,Huang2019,DJZhang2019} and recently have been successfully carried out with a single nitrogen-vacancy (NV) center in diamond \cite{DuJF2019,WLiu2020}. In these protocols the NH Hamiltonian emerges as an effective description of the subsystem for a larger dilated Hermitian system. We adopt this dilation formalism to simulate the NH nodal-line semimetal. In principle, the eigen-energies of NH Hamiltonian can be revealed from the dynamical phase, and the topological properties of the exceptional degeneracies can be reconstructed from the time-averaged spin textures.

\emph{Floquet-engineering of exceptional rings.--} We start by considering a model Hamiltonian exhibiting double exceptional rings,
\begin{equation}
\hat H_{\mathbf{k}}=\sum\nolimits_{\mathbf{k}}\hat \Psi_{\mathbf{k}}^\dagger \mathcal{H}(\mathbf{k}) \hat\Psi_{\mathbf{k}}+\sum\nolimits_{\mathbf{k},\sigma}(\hat c_{\mathbf{k},\sigma}\xi_{\mathbf{k},\sigma}^\dagger+\hat c_{\mathbf{k},\sigma}^\dagger \xi_{\mathbf{k},\sigma})\,,
\end{equation}
where $\hat \Psi_k=(\hat c_{\mathbf{k},a},\hat c_{\mathbf{k},b})^T$ and $a$, $b$ refer to the two internal degrees of freedom involved; $\mathcal{H}(\mathbf{k})=(m-Bk^2)\sigma_x+vk_z\sigma_z+i\gamma\sigma_z$ with $k^2=k_x^2+k_y^2+k_z^2$ and $\sigma_{x,y,z}$ are Pauli matrices; $v$ denotes the Fermi velocity along the $k_z$ direction, and $m$ and $B$ are system parameters with the dimension of energy and inverse energy, respectively \cite{Fang2015,ZhangHJ2019};  $\xi$ and $\xi^\dagger$ are the Langevin noise operators, and they satisfy $\langle \xi_{\mathbf{k},\sigma}(t)\xi_{\mathbf{k'},\sigma}^\dagger(t')\rangle_{\mathrm{noise}}=\pm2\gamma\delta_{\mathbf{k},\mathbf{k}'}\delta(t-t')$ for $\sigma=a,b$, $\langle \xi_{\mathbf{k},\sigma}^\dagger(t)\xi_{\mathbf{k}',\sigma}^\dagger(t')\rangle_{\mathrm{noise}}=0$  (see Appendix \ref{appa}) \cite{LPan2019}.

\begin{figure}[htbp]
	\centering
	\includegraphics[width=\textwidth]{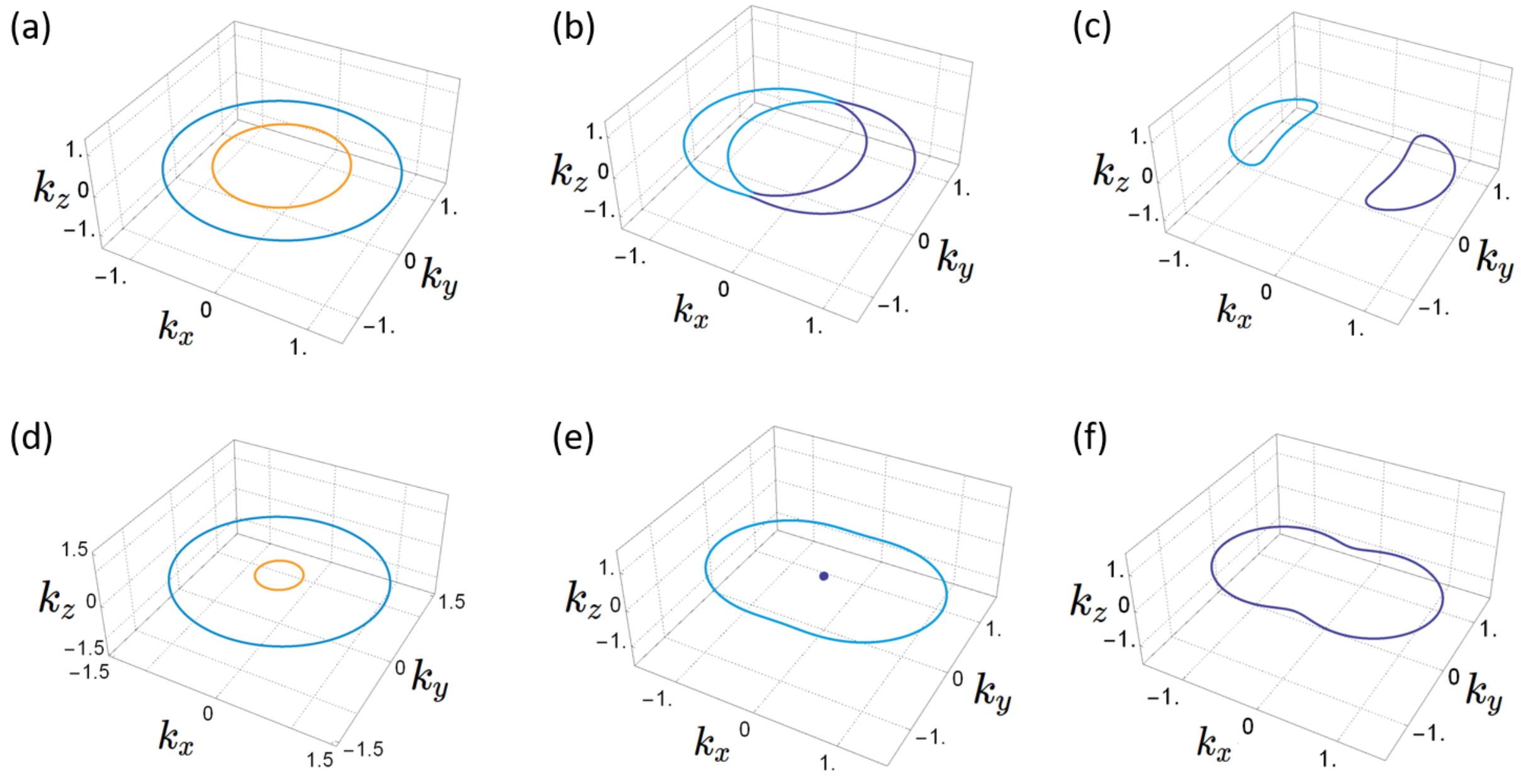}
	\caption{The band structure as computed from the Hamiltonian in Eq.(\ref{eq_flhm}) with $\tilde{m}=1$, $\phi=0$, (a-c) $\gamma=0.5$ and (d-f) $\gamma=0.9$; the frequency of the periodic driving (a) $\omega/A=10$, (b) $\omega/A=4.5$, (c) $\omega/A=3$, (d) $\omega/A=10$, (e) $\omega/A=3$, and (f) $\omega/A=2.5$. For simplicity we have taken $B=v=e=1$.}
	\label{fig1}
\end{figure}

We consider the periodic drive by application of a light beam which generates a vector potential $\mathbf{A}(t)=A[0,\cos(\omega t),\sin(\omega t+\phi)]$. The time-dependent Hamiltonian under the external field can be obtained by the minimal coupling prescription $\mathcal{H}(\mathbf{k}) \rightarrow \mathcal{H}(\mathbf{k}+e\mathbf{A})$. In the high frequency regime, the Floquet Hamiltonian can be computed in a perturbative manner. We follow the standard procedures to expand the prescribed Hamiltonian in orders of $1/\omega$ as $\mathcal{H}(t,\mathbf{k})=\sum_n\mathcal{H}_ne^{in\omega t}$,
\begin{equation}
\begin{aligned}
&\mathcal{H}_{0}(\mathbf{k}) =[m-\!B e^{2} A^{2}-\!B k^{2}] \Psi_{\mathbf{k}}^\dagger\sigma_{x}\Psi_{\mathbf{k}}+(v k_{z}\!+\!i\gamma) \Psi_{\mathbf{k}}^\dagger\sigma_{z}\Psi_{\mathbf{k}}\,, \\
&\mathcal{H}_{\pm 1}(\mathbf{k}) =-e A[2 B\left(k_{y} \mp i e^{\pm i \phi} k_{z}\right) \Psi_{\mathbf{k}}^\dagger\sigma_{x}\Psi_{\mathbf{k}}\\& \pm i e^{\pm i \phi} v \Psi_{\mathbf{k}}^\dagger\sigma_{z}\Psi_{\mathbf{k}}] / 2\,, \\
&\mathcal{H}_{\pm 2}(\mathbf{k}) =-B e^{2} A^{2}(1-e^{\pm i 2 \phi}) \Psi_{\mathbf{k}}^\dagger\sigma_{x}\Psi_{\mathbf{k}} / 4\,,
\end{aligned}
\end{equation}
and $\mathcal{H}_n=0$ for $|n|>2$. For the Floquet engineering of a non-Hermitian system, the effective time-independent Hamiltonian takes a slightly different form \cite{YChenup},
\begin{equation}
\begin{split}
\mathcal{H}_{\mathrm{eff}}(\mathbf{k})=&\mathcal{H}_{0}+\sum_{n \geq 1} \frac{[\mathcal{H}_{+n}, \mathcal{H}_{-n}]}{n \omega}+\\&\sum_{n,\sigma}s\frac{2i\gamma}{\omega^2}[\mathcal{H}_n,\hat c_{\mathbf{k},\sigma}]^\dagger[\mathcal{H}_n,\hat c_{\mathbf{k},\sigma}]\,,
\end{split}\label{eq_effh}
\end{equation}
where $s= 1(-1)$ for $\sigma=a(b)$, and the last term describes a micromotion. The first two terms lead to
\begin{equation}
\mathcal{H}_P=\left[\tilde{m}-B k^{2}\right] \sigma_{x}+(v k_{z}+i\gamma) \sigma_{z}+\lambda k_{y} \sigma_{y}\,,
\end{equation}
where $\lambda=-2e^2BvA^2\cos(\phi)/\omega$ and $\tilde{m}=m-Be^2A^2$, while the micromotion term gives rise to nontrivial second order contributions,
\begin{equation}
\begin{split}
&\sum\nolimits_{\sigma}[\mathcal{H}_1,\hat c_{\mathbf{k},\sigma}]^\dagger[\mathcal{H}_1,\hat c_{\mathbf{k},\sigma}]=\sum\nolimits_{\sigma}[\mathcal{H}_{-1},\hat c_{\mathbf{k},\sigma}]^\dagger[\mathcal{H}_{-1},\hat c_{\mathbf{k},\sigma}]\\
=&e^2A^2B^2(k_y^2+k_z^2+2k_yk_z\sin(\phi))\Psi_\mathbf{k}^\dagger\sigma_z\Psi_\mathbf{k}+\\
&\frac{e^2A^2v^2}{4}\Psi_\mathbf{k}^\dagger\sigma_z\Psi_\mathbf{k}-e^2A^2Bv(k_z+k_y\sin(\phi))\Psi_\mathbf{k}^\dagger\sigma_x\Psi_\mathbf{k}\,.
\end{split}
\end{equation}
Then the full form of the effective Hamiltonian is expressed as $\mathcal{H}_{\mathrm{eff}}(\mathbf{k})=\sum\nolimits_{\mathbf{k}}\hat \Psi_{\mathbf{k}}^\dagger \hat H_\mathbf{k}\hat\Psi_{\mathbf{k}}$ with
\begin{equation}
\begin{split}
\hat H_k=&[\tilde{m}-Bk^2-i(\gamma_1k_y\sin(\phi)+\gamma_1k_z)]\sigma_x+\lambda k_y\sigma_y\\+&[vk_z+i(\gamma_2-\gamma_3(k_y^2+k_z^2+2k_yk_z\sin\phi))]\sigma_z\,,
\end{split}\label{eq_flhm}
\end{equation}
where $\gamma_1=4e^2A^2Bv\gamma/\omega^2$ and $\gamma_2=\gamma(1+e^2A^2v^2/\omega^2)$ and $\gamma_3=4e^2A^2B^2\gamma/\omega^2$. We can rewrite the Hamiltonian in a more compact form,
\begin{equation}
\hat H_k=\mathbf{d}(\mathbf{k}) \cdot \boldsymbol{\sigma}, \quad \mathbf{d} \in \mathbb{C}^{3}\,,\label{eq_gnh}
\end{equation}
where the vector $\mathbf{d}$ can be decomposed into real and imaginary parts according to $\mathbf{d}=\mathbf{d}_R+i\mathbf{d}_I$. Then the eigenvalues and thus energies take a general form,
\begin{equation}
\epsilon_{\pm}^{2}=\mathbf{d}_{R}^{2}-\mathbf{d}_{I}^{2}+2 i \mathbf{d}_{R} \cdot \mathbf{d}_{I}\,.
\end{equation}
For the non-Hermitian system we consider, $\epsilon_\pm=\pm\sqrt{A(\mathbf{k})}e^{i\theta/2}$, where $A(\mathbf{k})=\sqrt{(\mathbf{d}_{R}^{2}-\mathbf{d}_{I}^{2})^2+4(\mathbf{d}_{R} \cdot \mathbf{d}_{I})^2}$ and $\theta=\arctan[2(\mathbf{d}_{R} \cdot \mathbf{d}_{I})/(\mathbf{d}_{R}^{2}-\mathbf{d}_{I}^{2})]$. Thus the eigenvalues of Hamiltonian possess two branches and then acquire a nontrivial energy vorticity.

We calculate the band structure from the Floquet Hamiltonian defined in Eq. (\ref{eq_flhm}). The results are shown in Fig. \ref{fig1}. The nodal points in the spectrum are described by the conditions,
\begin{equation}
\mathbf{d}_{R}^{2}-\mathbf{d}_{I}^{2}=0\,,\quad \mathbf{d}_{R} \cdot \mathbf{d}_{I}=0\,.
\end{equation}
In the high-frequency limit $\omega \gg 1$, the term dependent on $\omega$ can be neglected and the energy spectrum is approximated by,
\begin{equation}
\epsilon_{\pm}=\pm\sqrt{(\tilde m-Bk^2)^2+v^2k_z^2-\gamma^2+2ivk_z\gamma}\,.
\end{equation}
When $\gamma < \tilde{m}$, two ERs characterized by $Bk^2=\tilde{m}\pm\gamma$ lie in the $k_z=0$ plane, as shown in Fig. \ref{fig1} (a). The inner ER shrinks as $\gamma$ increases, and vanishes beyond a critical value of $\gamma = \tilde{m}$ where it becomes an EP. As the driving frequency $\omega$ decreases and the system goes into the lower-frequency regime, the structure of the exceptional solutions dramatically changes. We consider $\gamma=0.5\tilde{m}$ as a typical example [For a comparison between quasienergy spectrum of the full Floquet Hamiltonian and that of the effective Hamiltonian (in Hermitian limit), see Appendix \ref{app_fullspc}.]. The inner ER extends and cuts the outer ER to form an unconcentric dipolar pair, leading to a nontrivial Lifshitz transition of the Fermi surface, as illustrated in Fig. \ref{fig1} (c).

In particular, for the critical non-Hermiticity $\gamma=\tilde{m}$, an ER accompanied with a single EP appears in the spectrum in the high-frequency limit. For smaller $\gamma \sim \tilde{m}$, it is also possible to drive the inner ER into a single EP, and then even lift it , as shown in Figs. \ref{fig1}(d)-\ref{fig1}(e).

\begin{figure}[htbp]
	\centering
	\includegraphics[width=\textwidth]{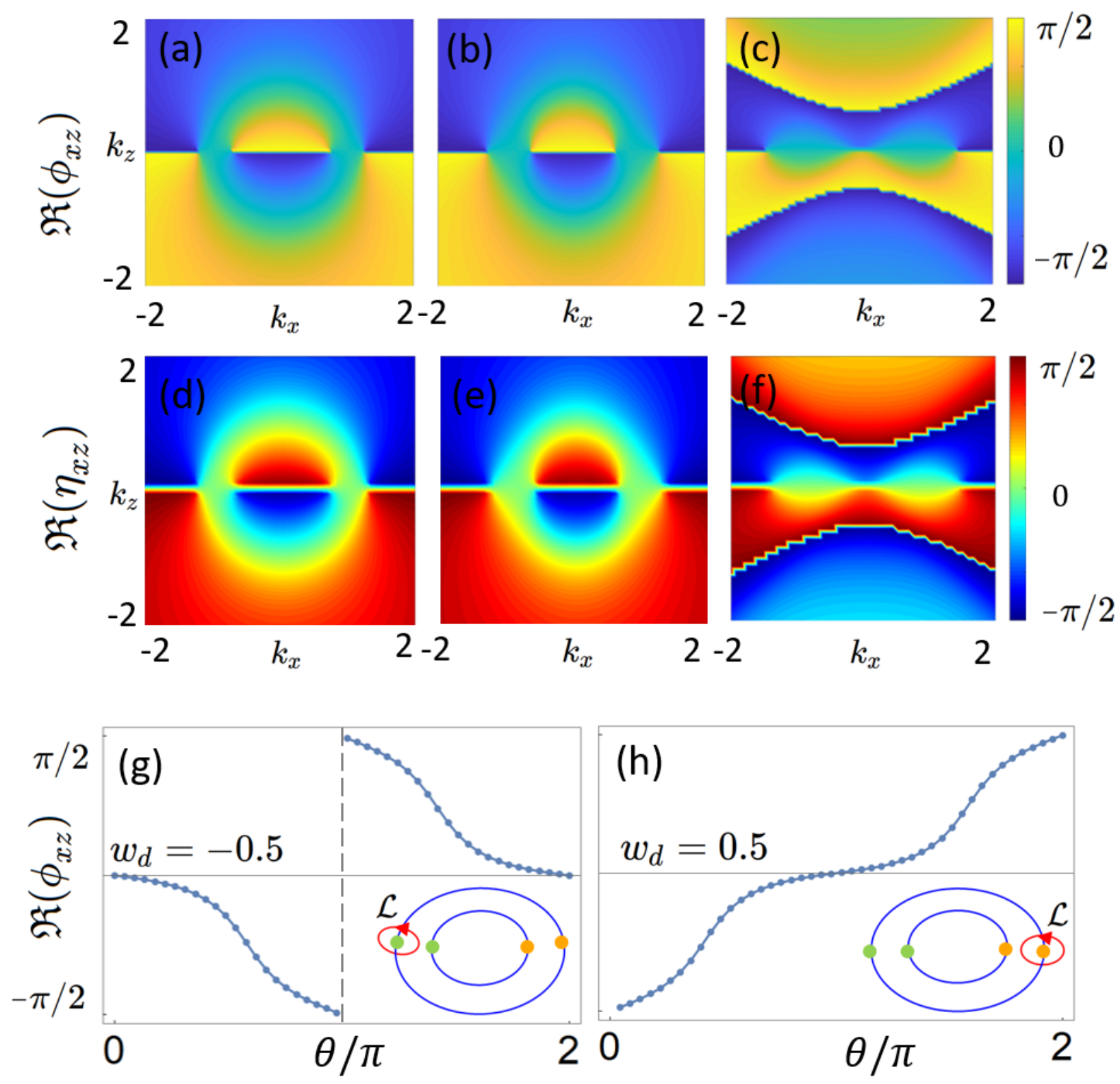}
	\caption{Panels (a), (b), and (c), respectively, show the relative phase $\phi_{xz}(k_x,k_z)$ with $\tilde{m}=1$, $B=1$, $v=1$, $\gamma=0.5$ and (a) $\omega/A=10$, (b) $\omega/A=2$ and (c) $\omega/A=1$; panels (d), (e), and (f), respectively, show the relative phase $\eta_{xz}(k_x,k_z)$ of spin textures for an evolution time $T=20$; panels (g) and (h) show the winding number extracted from the loops encircling the two exceptional points on the inner ER.}
	\label{fig2}
\end{figure}

\emph{Topological characterization--}The exceptional rings and exceptional points are topological defects which can be characterized by a quantized Berry phase,
\begin{equation}
\gamma_B=\oint_{2\mathcal{L}} i\langle \tilde{u}(\mathbf{k})|\partial_{\mathbf{k}}u(\mathbf{k})\rangle d\mathbf{k}\,,
\end{equation}
where $\langle\tilde u(\mathbf{k})|$ and $|u(\mathbf{k})\rangle$ are left and right eigenvectors of the lower band of the Hamiltonian (\ref{eq_flhm}) respectively; the path $2\mathcal{L}$ travels across the ring twice. The path $2\mathcal{L}$ forms a closed loop on the Riemann surface defined by $\epsilon(\theta)$ \cite{XuDuan2017}, and the system returns to its original state after wrapping the ER twice.

Without loss of generality, we consider a path in the $k_y=0$ plane. For the two-band model we consider, the Berry phase can be associated to a winding number with the relation $\gamma_B=2\pi w$, where
\begin{equation}
w=\frac{1}{2\pi}\oint_\mathcal{L} \partial_\mathbf{k} \phi_{xz}(\mathbf{k})~d\mathbf{k}\,,\label{eq_wn}
\end{equation}
with $\phi_{xz}(\mathbf{k})\equiv\arctan(h_x/h_z)$. The winding number can be experimentally detected with a dynamic approach from the long-time average of spin textures \cite{LeeCH2019,LZhou2019}. The spin textures are defined as the expectation values of the Pauli matrices $\langle\sigma_j(\mathbf{k},t)\rangle=\langle\tilde{u}_\mathbf{k}|\sigma_j|u_\mathbf{k}\rangle/\langle\tilde{u}_\mathbf{k}|u_\mathbf{k}\rangle$ in a biorthogonal formalism. The dynamic winding number is defined by the spin vectors
\begin{equation}
w_d=\frac{1}{2\pi}\oint_{\mathcal{L}} \partial_\mathbf{k} \eta_{ij}(\mathbf{k})d\mathbf{k}\,,
\end{equation}
where $\eta_{ij}(\mathbf{k})\equiv \arctan(\overline{\sigma}_j/\overline{\sigma}_i)$ and $\overline{\sigma}_j=\frac{1}{T}\int_0^T\langle\sigma_j(\mathbf{k},t)\rangle dt$. In the long-time limit, this dynamic winding number is equivalent to the winding number $w=\lim_{T \to \infty} w_d$. Only the real part of the phase $\eta_{ij}(\mathbf{k})$ has nontrivial contributions and it can be decomposed to the sum of two observables \cite{CYin2018},
\begin{equation}
\Re(\eta_{ij}(\mathbf{k}))=\frac{1}{2}(\phi_{ij}^{RR}+\phi_{ij}^{LL})+n\frac{\pi}{2}\,,
\end{equation}
where $\phi_{ij}^{RR}=\arctan(\overline{\langle u|\sigma_i|u\rangle}/\overline{\langle u|\sigma_j|u\rangle})$ and $\phi_{ij}^{LL}=\arctan(\overline{\langle \tilde{u}|\sigma_i|\tilde{u}\rangle}/\overline{\langle \tilde{u}|\sigma_j|\tilde{u}\rangle})$.

As shown in Fig. \ref{fig2}, the ERs are characterized by the winding number $w=0.5$ along an $S^1$ loop which interlinks with them [the sign difference between the winding numbers in Figs. \ref{fig2}(g) and \ref{fig2}(f) can be canceled by orienting the integral path in a right-hand rule] \cite{ZYang2020}. The fact that the winding number takes value out of $\mathbb{Z}/2$ can be attributed to the net vorticity of this NH system. Furthermore, numerical simulations show that the dynamical winding number has good agreement with the winding number, as shown in Figs. \ref{fig2}(d)-\ref{fig2}(f).

\begin{figure}[htbp]
	\centering
	\includegraphics[width=\textwidth]{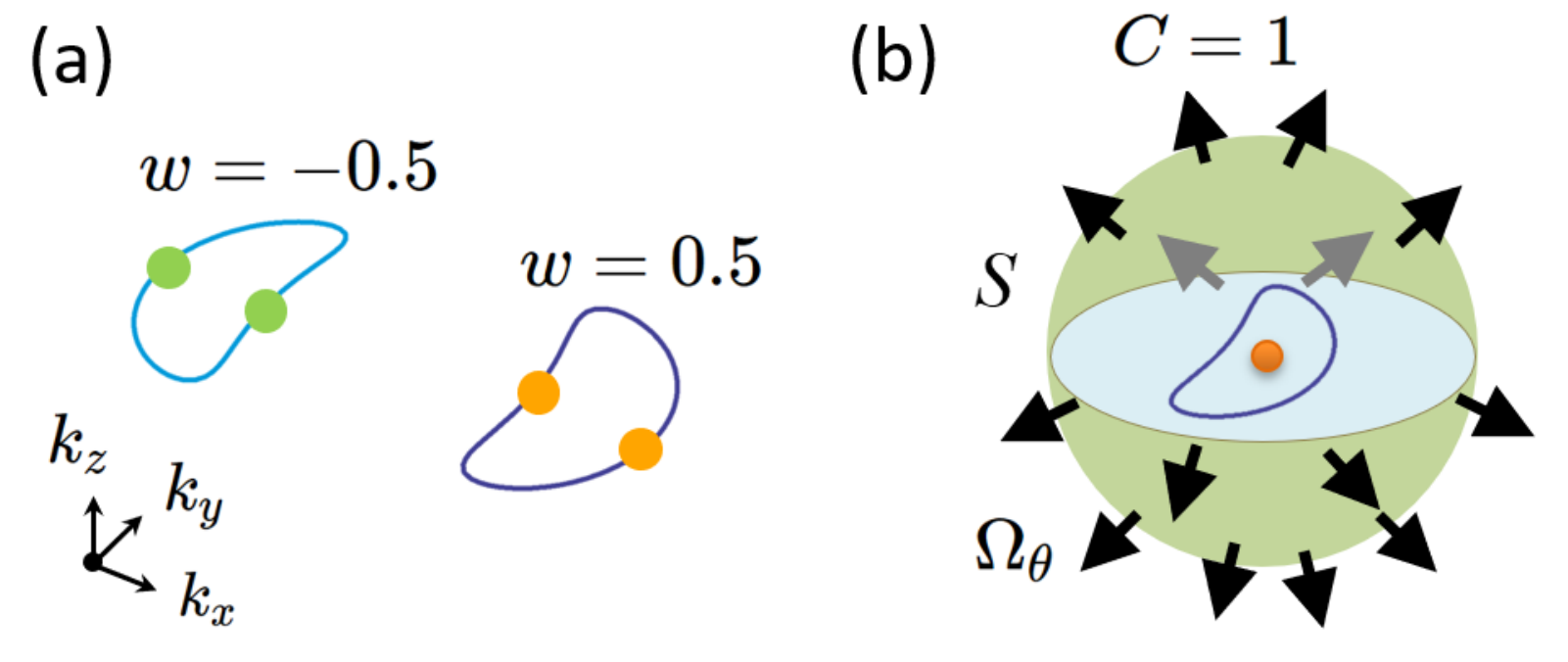}
	\caption{(a) The winding number of the ER, which is $0.5$ for an integral loop which interlinks with the ER, and 1 for that encircling the ER, (b) The Chern number of the ER. The ER is analogous to the Weyl point in the Hermitian case, indicated with a colored dot in (b).}
	\label{fig3}
\end{figure}

When the system goes into the lower-frequency regime, the double concentric ERs are deformed as a dipolar pair with two pole ERs which carry opposite topological charges $w=\pm 1$ defined on the loop encircling the whole ER [indicated in Fig. \ref{fig3} (a)]. The net winding number implies that the ER is protected by a Chern number (see Appendix \ref{app_chern} for details),
\begin{equation}
C=\frac{1}{2 \pi} \oint_{\mathcal{S}} \mathbf{\Omega}_{\theta}(\boldsymbol{k}) \cdot d \boldsymbol{S}\,,
\end{equation}
where $\boldsymbol{\Omega}_{\theta}(\boldsymbol{k})=i\langle\nabla_{\boldsymbol{k}} \tilde{u}_{\theta}(\boldsymbol{k})|\times| \nabla_{\boldsymbol{k}} u_{\theta}(\boldsymbol{k})\rangle$ is the Berry curvature and $\mathcal{S}$ encloses whole single ER, as illustrated in Fig. \ref{fig3} (b). Numerical calculation shows that $C=\pm 1$.

\emph{Hall conductivity.--}The light beam is incident along the $x$ direction, thus we are interested in the perpendicular $yz$-component of the Hall response $\sigma_{yz}$. As suggested by previous works \cite{Hirsbrunner2019,HZhai2018}, the Hall conductivity can be calculated following from the linear response theory,
\begin{equation}
\sigma_{yz}=\lim _{\omega \rightarrow 0} \frac{i}{\omega+i 0^{+}}(K_{yz}(\omega)-K_{zy}(0))\,,
\end{equation}
where $K_{yz}$ is the current-current correlation function (for details, see Appendix \ref{Appf}).

For a generic two-band Hamiltonian with the form of Eq. (\ref{eq_gnh}) at zero temperature ($\beta=\infty$), the Hall conductivity is given by \cite{HZhai2018},
\begin{equation}
\sigma_{yz}=\sum_{\mathbf{k}} \frac{\Omega_{yz}(\mathbf{k})+\Omega_{yz}^{*}(\mathbf{k})}{2} \times v_{\mathbf{k}}\,,
\end{equation}
where $\Omega_{yz}(\mathbf{k})=\mathbf{d}({\mathbf{k}}) \cdot (\partial_{k_{y}} \mathbf{d}_R({\mathbf{k}}) \times \partial_{k_{z}} \mathbf{d}_R({\mathbf{k}}) ) / \epsilon(\mathbf{k})^{3}$ and $v_\mathbf{k}=2\arctan[\Re \epsilon(\mathbf{k})/|\Im \epsilon(\mathbf{k})|]/\pi$.

\begin{figure}[htbp]
	\centering
	\includegraphics[width=0.8\textwidth]{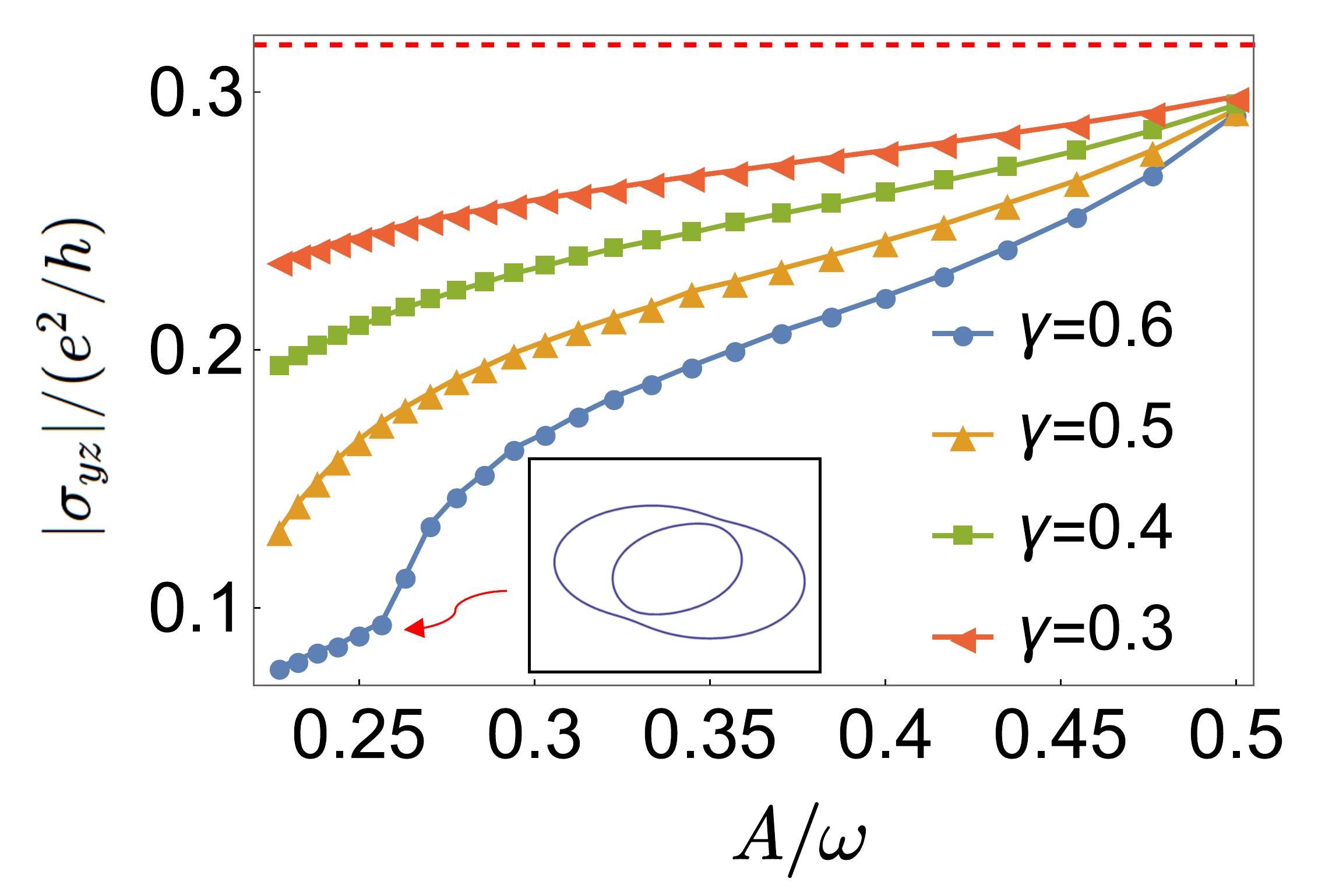}
	\caption{The dependence of Hall conductivity on the frequency $A/\omega$ of periodic driving. The parameters are $\tilde{m}=1$, $B=1$, $v=1$, $\phi=0$. The dashed line shows the value of $\sigma_{yz}$ in the Hermitian limit $\sigma_0=e^2/\pi h\sqrt{\tilde{m}/B}$. The inset shows the exceptional degeneracies at the critical point for $\gamma=0.6$.}
	\label{fig4}
\end{figure}

We numerically calculate the integral, the results are shown in Fig. \ref{fig4}. In general, the intimate relation between the Chern number and the Hall conductivity does not hold for NH Hamiltonians. With the inclusion of a self-energy term describing the non-Hermiticity, the Green function becomes discontinuous in the $(\omega,\mathbf{k})$ space, thus not a homeomorphism \cite{Hirsbrunner2019}. The Hall conductivity is not quantized, however, and still signals the topological transition. In the Hermitian limit, $\sigma_{yz}=e^2/(\pi\hbar)\sqrt{m/B-e^2A^2}$, which is  proportional to the distance between the two Weyl points, but with additional first-order contribution $\sim \sqrt{A^2}$ comparing to the conventional Weyl semimetals \cite{Wang2016}. With nonzero $\gamma$, the criticality of the Hall conductivity is observed across a critical point where the dipolar ER pair forms, distinguishing from the Hermitian case where this transition immediately happens under the driven field. Furthermore, the Hall conductivity decreases as the strength of the non-Hermiticity increases since high-order effect induced by the NH terms becomes more dominant.

\emph{Quantum simulation with a quantum spin system.--}To simulate the dynamics of the Hamiltonian $\hat H_\mathbf{k}$, an ancilla qubit is required to dilate $\hat H_\mathbf{k}$ into a Hermitian Hamiltonian $\hat H_d(t)$ \cite{DuJF2019}. The evolution of the dilated system $\hat H_d(t)$ is governed by the Schr$\mathrm{\ddot{o}}$dinger equation,
\begin{equation}
i \frac{d}{d t}|\Psi(t)\rangle= \hat H_{d}(t)|\Psi(t)\rangle\,,
\end{equation}
where $|\Psi(t)\rangle$ is the state of the combined system. The essential idea that allows the exclusive dilation is to restrict the measurement results to those with a specific output for the ancilla qubit. In such a postselection scenario, it's convenient to write the state $|\Psi(t)\rangle$ in a form
\begin{equation}
|\Psi(t)\rangle=|\psi(t)\rangle|-\rangle_a+\eta(t)|\psi(t)\rangle|+\rangle_a\,,
\end{equation}
where $|-\rangle_a$ and $|+\rangle_a$ form an orthonormal basis of the ancilla qubit  and here are chosen to be the eigenstates of $\sigma_y$ with $|-\rangle=(|0\rangle- i|1\rangle) / \sqrt{2}$ and $|+\rangle=-i(|0\rangle+i|1\rangle) / \sqrt{2}$. After the evolution, a $-\pi/2$ pulse is applied and only the measurement results with no jump outside the submanifold $|\psi\rangle|1\rangle_a$ are postselected.

\begin{figure}[htbp]
	\centering
	\includegraphics[width=\textwidth]{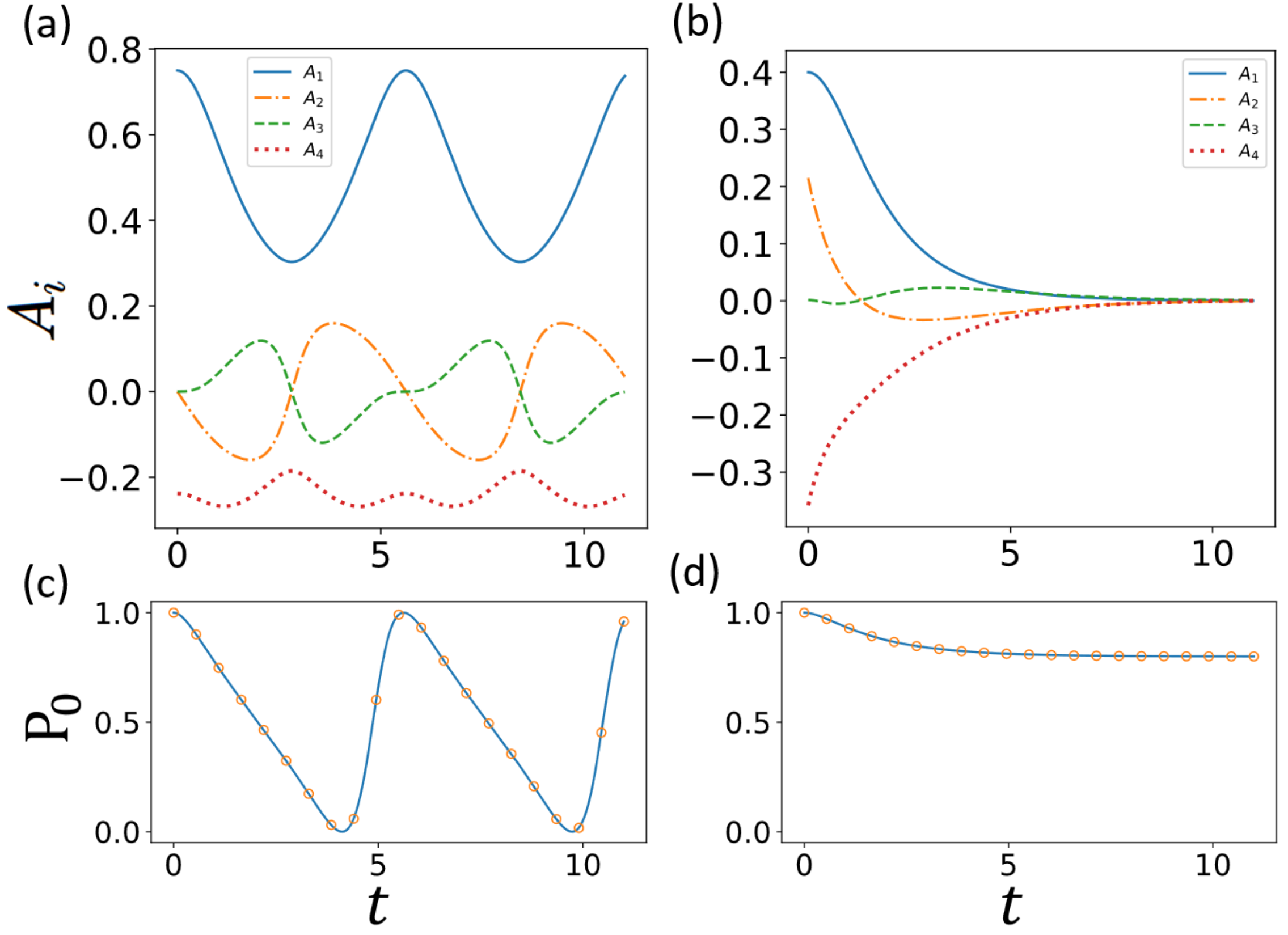}
	\caption{(a, b) Parameters $A_i(t)$ in the dilated Hamiltonian. In general $B_i's \neq 0$, but for parameters considered here all $B_i's=0$. (c, d) The population on state $|0\rangle$ at time $t$. The solid lines indicate the analytic results solved for the Schr$\rm{\ddot{o}}$dinger equation $i\hbar\partial_t \psi=\hat H_\mathbf{k} \psi$; the circles indicate the numerical results solved for dilated Hamiltonian $\hat H_d$ after postselection.  The parameters in $\hat H_\mathbf{k}$ are chosen as $\tilde{m}=v=B=1$ and (a,c) $k_x^2+k_y^2=0.25$, $k_z=0$, (b,d) $k_x^2+k_y^2=0.6$, $k_z=0$.}
	\label{fig5}
\end{figure}

Note that the Hamiltonian $\hat H_d(t)$ is not uniquely determined. One proper choice is
\begin{equation}
\hat H_{d}(t)=\Lambda(t) \otimes \mathbb{I}+\Gamma(t) \otimes \sigma_{z}\,,
\end{equation}
where $\Lambda(t)=\{\hat H_{\mathbf{k}}(t)+[i \frac{d}{d t} \eta(t)+\eta(t) \hat H_{\mathbf{k}}(t)] \eta(t)\} M^{-1}(t)$ and $\Gamma(t)=i[\hat H_{\mathbf{k}}(t) \eta(t)-\eta(t) \hat H_{\mathbf{k}}(t)-i \frac{d}{d t} \eta(t)] M^{-1}(t)$. The time-dependent operator $M(t)$ takes the form $M(t)=\eta^\dagger\eta+I$. We can expand $\Lambda(t)$ and $\Gamma(t)$ in terms of the Pauli operators and rewrite $\hat H_d(t)$ as
\begin{equation}
\begin{split}
\hat H_d=&A_{1}(t) \sigma_{x} \!\otimes\! \mathbb{I}+\! A_{2}(t) \mathbb{I} \!\otimes\! \sigma_{z}+\! A_{3}(t) \sigma_{y} \!\otimes\! \sigma_{z}+\!  A_{4}(t) \sigma_{z} \!\otimes\! \sigma_{z}\\
+&B_{1}(t) \mathbb{I} \!\otimes\! \mathbb{I}+B_{2}(t) \sigma_{y} \!\otimes\! \mathbb{I}+B_{3}(t) \sigma_{z} \!\otimes\! \mathbb{I}+B_{4}(t) \sigma_{x} \!\otimes\! \sigma_z\,.
\end{split}\label{eq_hd}
\end{equation}
Figures \ref{fig5}(a) and \ref{fig5}(b) show the time-dependent parameters $A_i$ ($i=$1-4). Without loss of generality, here we take only the Hamiltonian in the high-frequency limit as an example. The four-level system described by the Hamiltonian $\hat H_d$ could be encoded in the ground state manifold of the electron spin and the nuclear spin in a NV center, or alternatively, other  quantum platforms such as trapped ions, Rydberg atoms, and superconducting circuits. As shown in Fig. \ref{fig5} (c) and (d), we demonstrate that the dynamics of the NH nodal-line semimetal can be revealed in the post-selected state population, which allows us to reconstruct the topological information from the spin textures. In experiments, microwave pulses and two radio-frequency pulses could be applied to couple the ground states of a NV center. For the coupling strength tuned to $\sim600~\rm{kHz}$, the dynamical process in Figs. \ref{fig5}(c) and \ref{fig5}(d) can happen within $10~\rm{\mu s}$, which is feasible with current technology.

Finally, we would like to address another possible scheme realizing the NH nodal-line semimetal with dissipative ultracold atomic systems. The two-band nodal-line semimetal was recently implemented with ultracold fermions using Raman-induced spin-orbit coupling \cite{BSong2019}. To introduce non-Hermiticity into the lattice, on-site atom loss could be generated by applying a radio frequency pulse to resonantly transfer the atoms to an irrelevant excited state \cite{JLi2019,XuDuan2017}.

\emph{Discussions and conclusions.--}The perturbation with on site gain and loss preserves the mirror symmetry of the Hamiltonian, $\mathcal{M}\mathcal{H}(k_{x,y},k_z)\mathcal{M}^{-1}=\mathcal{H}(k_{x,y},-k_z)$, with $\mathcal{M}=i\sigma_x$, but breaks the $\mathcal{PT}$-symmetry. For a $\mathcal{PT}$-symmetric NH perturbation such as $i\gamma_y\sigma_y$ which describes nonreciprocal hooping, the exceptional structure even forms a surface. This exceptional surface could be driven into a pair of ERs in the $xz$-plane by the light (here for simplicity we do not consider the Langevin noise). Although not a focus here, the Floquet driving of NH systems with new symmetry, is an interesting direction for further exploration.

In summary, we have proposed a scheme to realize tunable exceptional rings in terms of Floquet engineering. As the driven frequency changes, a dipolar pair of ERs protected by opposite Chern numbers can be created from the NH semimetal with double concentric ERs. This transition process is accompanied by the criticality of the Hall conductivity. Furthermore, we explore possible realization with synthetic quantum matter, which does not require controlling an open system. The proposed system would provide a promising platform for elaborating NH topology which might be elusive in nature.

\emph{Note added.--}Recently we became aware of a related eprint by A. Banerjee and A. Narayan \cite{Banerjee2020}.

\vspace{0.5cm}
\begin{acknowledgments}
The authors thank S. L. Zhu for useful discussions.
The work was supported by  the National Natural Science Foundation of China (Grants No. 91636218 and No. U1801661) and the National Key Research and Development Program of China (Grants No. 2016YFA0301803).
\end{acknowledgments}

\begin{appendix}
\section{The Langevin noise and Floquet Hamiltonian}\label{appa}
In this section, we elaborate the role of the Langevin noise, following the theory developed by Pan \emph{et al.} \cite{LPan2019,YChenup}. The Langevin noise describes the coupling between the open system and the environment. When one integrates out the environment to obtain a NH Hamiltonian, it always comes together with the Langevin noise. The Langevin noise conserves the trace of the density matrix and ensures the unitarity of the quantum operators. The time evolution operator can be written in a time-ordered integral form as
\begin{equation}
\mathcal{U}(t)=\mathcal{T}[\exp(-i\int_0^t \hat H_{\mathbf{k}}(t') dt')]\,,
\end{equation}
where $\mathcal{T}$ is time-ordering operator. One can check that in the linear response level and after noise average, the time evolution operator is unitary \cite{LPan2019},
\begin{equation}
\begin{split}
\mathcal{U}^\dagger(t)\mathcal{U}(t)=&\mathbb{I}+\gamma\int_0^t dt' \sum_{\mathbf{k}}2\Psi_{\mathbf{k}}^\dagger\sigma_z\Psi_{\mathbf{k}}+\\&\sum_{\mathbf{k},\mathbf{k}',\sigma}\int \int dt' dt''\langle \hat c_{\mathbf{k},\sigma}^\dagger(t')\hat c_{\mathbf{k'},\sigma}(t'')\xi_{\mathbf{k}}(t')\xi^\dagger_{\mathbf{k}'}(t'')\rangle_{\mathrm{noise}}\\
=&\mathbb{I}+2\gamma\int_0^t dt'\sum_{\mathbf{k},\sigma}(\hat c_{\mathbf{k},\sigma}^\dagger(t')\hat c_{\mathbf{k},\sigma}(t')-\hat c_{\mathbf{k},\sigma}^\dagger(t')\hat c_{\mathbf{k},\sigma}(t'))\\
=&\mathbb{I}\,.
\end{split}
\end{equation}
Then we consider the time evolution of any operator $\hat W_H(t)=\mathcal{U}^\dagger(t) \hat W \mathcal{U}(t)$. Expanded to the first order,
\begin{equation}
\begin{split}
\hat W_H(t)=&\hat W(t)+\gamma\int_{0}^{t} dt' \sum_{\mathbf{k},\sigma}\Big(2s \hat c_{\mathbf{k},\sigma}^\dagger(t')\hat W(t)\hat c_{\mathbf{k},\sigma}^\dagger(t')\\
&-s\{\hat W(t),\hat c_{\mathbf{k},\sigma}^\dagger(t')\hat c_{\mathbf{k},\sigma}(t'))\}\Big)\,,
\end{split}
\end{equation}
where $\hat W(t)=e^{i\hat H_0t}\hat W e^{-i\hat H_0t}$. The physical observable W is given by $\mathcal{W}=\langle {\rm{Tr}}(\rho_0\hat W_H(t))\rangle$, where $\rho_0=e^{-\beta H_0}/{\rm{Tr}}(e^{-\beta H_0})$ is the initial equilibrium density matrix of the nonperturbed Hermitian system. Upon the first order approximation, the evolution of the observable has an effective Hamiltonian representation, which gives rise to Eq. (\ref{eq_effh}) with series expansion.

We remark that the Langevin noise is essential in the present formalism. The open system with particle gain and loss can also be described by the Lindblad master equation in an alternative way. However, this does not necessitate an effective Hamiltonian description. The system hosts an effective Hamiltonian only when the quantum jump term is negligible. That means the Hamiltonian in the absence of the Langevin noise describes the dynamics well only over a short time \cite{Yamamoto2019}. Therefore, for the long-time Floquet evolution considered here, the NH Hamiltonian together with the Langevin noise provide a complete description.

\section{Derivation details of Eq.(5)}
In this section, we give the detailed derivation of the micromotion term of Eq. (\ref{eq_effh}). We recall the full form of the first-order component of the Floquet Hamiltonian,
\begin{equation}
\mathcal{H}_{+1}=-\Psi_\mathbf{k}^\dagger[eAB(k_y-i e^{i\phi})\sigma_x+i \frac{evA}{2}e^{i\phi}\sigma_z]\Psi_\mathbf{k}\,,
\end{equation}
\begin{equation}
\mathcal{H}_{-1}=-\Psi_\mathbf{k}^\dagger[eAB(k_y+i e^{-i\phi})\tau_x-i \frac{evA}{2}e^{-i\phi}\tau_z]\Psi_\mathbf{k}\,,
\end{equation}
expanded as
\begin{equation}
\begin{split}
\mathcal{H}_{+1}=-[&eAB(k_y-i e^{i\phi})(\hat c_{\mathbf{k},a}^\dagger  \hat c_{\mathbf{k},b}+\hat c_{\mathbf{k},b}^\dagger  \hat c_{\mathbf{k},a})+\\&i \frac{evA}{2}e^{i\phi}(\hat c_{\mathbf{k},a}^\dagger  \hat c_{\mathbf{k},a}-\hat c_{\mathbf{k},b}^\dagger  \hat c_{\mathbf{k},b})]\,,
\end{split}
\end{equation}
\begin{equation}
\begin{split}
\mathcal{H}_{-1}=-[&eAB(k_y+i e^{-i\phi})(\hat c_{\mathbf{k},a}^\dagger  \hat c_{\mathbf{k},b}+\hat c_{\mathbf{k},b}^\dagger  \hat c_{\mathbf{k},a})-\\&i \frac{evA}{2}e^{-i\phi}(\hat c_{\mathbf{k},a}^\dagger  \hat c_{\mathbf{k},a}-\hat c_{\mathbf{k},b}^\dagger  \hat c_{\mathbf{k},b})]\,.
\end{split}
\end{equation}
The commutators are given by
\begin{equation}
[\mathcal{H}_{+1},\hat c_{\mathbf{k},a}]=eAB(k_y-i k_ze^{i\phi})\hat c_{\mathbf{k},b}+i \frac{evA}{2}e^{i\phi}\hat c_{\mathbf{k},a}\,,\label{eq_c1}
\end{equation}
\begin{equation}
[\mathcal{H}_{+1},\hat c_{\mathbf{k},a}]^\dagger=eAB(k_y+i k_z e^{-i\phi})\hat c_{\mathbf{k},b}^\dagger-i \frac{evA}{2}e^{-i\phi}\hat c_{\mathbf{k},a}^\dagger\,.\label{eq_c2}
\end{equation}
\begin{equation}
[\mathcal{H}_{+1},\hat c_{\mathbf{k},b}]=eAB(k_y-i k_z e^{i\phi})\hat c_{\mathbf{k},a}-i \frac{evA}{2}e^{i\phi}\hat c_{\mathbf{k},b}\,,\label{eq_c3}
\end{equation}
\begin{equation}
[\mathcal{H}_{+1},\hat c_{\mathbf{k},b}]^\dagger=eAB(k_y+i e^{-i\phi})\hat c_{\mathbf{k},a}^\dagger+i \frac{evA}{2}e^{-i\phi}\hat c_{\mathbf{k},b}^\dagger\,.\label{eq_c4}
\end{equation}
By inserting Eq. (\ref{eq_c1}) and (\ref{eq_c2}) into Eq. (\ref{eq_effh}), we have
\begin{equation}
\begin{split}
&\frac{2i\gamma}{\omega^2}[\mathcal{H}_{+1},\hat c_{\mathbf{k},a}]^\dagger[\mathcal{H}_{+1},\hat c_{\mathbf{k},a}]=\\&\frac{2i\gamma}{\omega^2}\{e^2A^2B^2[k_y^2+k_z^2-2k_y k_z\sin(\phi)]\hat c_{\mathbf{k},b}^\dagger\hat c_{\mathbf{k},b}+\\
&\frac{e^2A^2Bv}{2}(i k_ye^{i\phi}-k_z)\hat c_{\mathbf{k},b}^\dagger\hat c_{\mathbf{k},a}+\frac{e^2A^2v^2}{4}\hat c_{\mathbf{k},a}^\dagger\hat c_{\mathbf{k},a}-\\
&\frac{e^2A^2Bv}{2}(i k_ye^{-i\phi}+k_z)\hat c_{\mathbf{k},a}^\dagger\hat c_{\mathbf{k},b}\}\,.
\end{split}\label{eq_mcm1}
\end{equation}
By inserting Eq. (\ref{eq_c3}) and (\ref{eq_c4}) into Eq. (\ref{eq_effh}), we have
\begin{equation}
\begin{split}
&-\frac{2i\gamma}{\omega^2}[\mathcal{H}_{+1},\hat c_{\mathbf{k},b}]^\dagger[\mathcal{H}_{+1},\hat c_{\mathbf{k},b}]=\\&-\frac{2i\gamma}{\omega^2}\{e^2A^2B^2[k_y^2+k_z^2-2k_y k_z\sin(\phi)]\hat c_{\mathbf{k},a}^\dagger\hat c_{\mathbf{k},a}-\\
&\frac{e^2A^2Bv}{2}(i k_ye^{i\phi}-k_z)\hat c_{\mathbf{k},a}^\dagger\hat c_{\mathbf{k},b}+\frac{e^2A^2v^2}{4}\hat c_{\mathbf{k},b}^\dagger\hat c_{\mathbf{k},b}+\\
&\frac{e^2A^2Bv}{2}(i k_ye^{-i\phi}+k_z)\hat c_{\mathbf{k},b}^\dagger\hat c_{\mathbf{k},a}\}\,.
\end{split}\label{eq_mcm2}
\end{equation}
Collecting all terms in Eq. (\ref{eq_mcm1}) and Eq. (\ref{eq_mcm2}) gives
\begin{equation}
\begin{split}
&\sum_{\sigma}s\frac{2i\gamma}{\omega^2}[\mathcal{H}_{+1},\hat c_{\mathbf{k},\sigma}]^\dagger[\mathcal{H}_{+1},\hat c_{\mathbf{k},\sigma}]=\\
&-\frac{2i\gamma}{\omega^2}[e^2A^2B^2(k_y^2+k_z^2-2k_y k_z\sin(\phi))-\frac{e^2A^2v^2}{4}]\Psi_\mathbf{k}^\dagger\sigma_z\Psi_\mathbf{k}\\
&-\frac{2i\gamma}{\omega^2}\frac{e^2A^2Bv}{2} [2k_y\sin(\phi)+2k_z] \Psi_\mathbf{k}^\dagger\sigma_x\Psi_\mathbf{k}\,.
\end{split}
\end{equation}
In the same way we can also obtain that,
\begin{equation}
[\mathcal{H}_{-1},\hat c_{\mathbf{k},\sigma}]^\dagger[\mathcal{H}_{-1},\hat c_{\mathbf{k},\sigma}]=[\mathcal{H}_{+1},\hat c_{\mathbf{k},\sigma}]^\dagger[\mathcal{H}_{+1},\hat c_{\mathbf{k},\sigma}]\,.
\end{equation}

\section{Energy dispersion}

\begin{figure}[htbp]
	\centering
	\includegraphics[width=\textwidth]{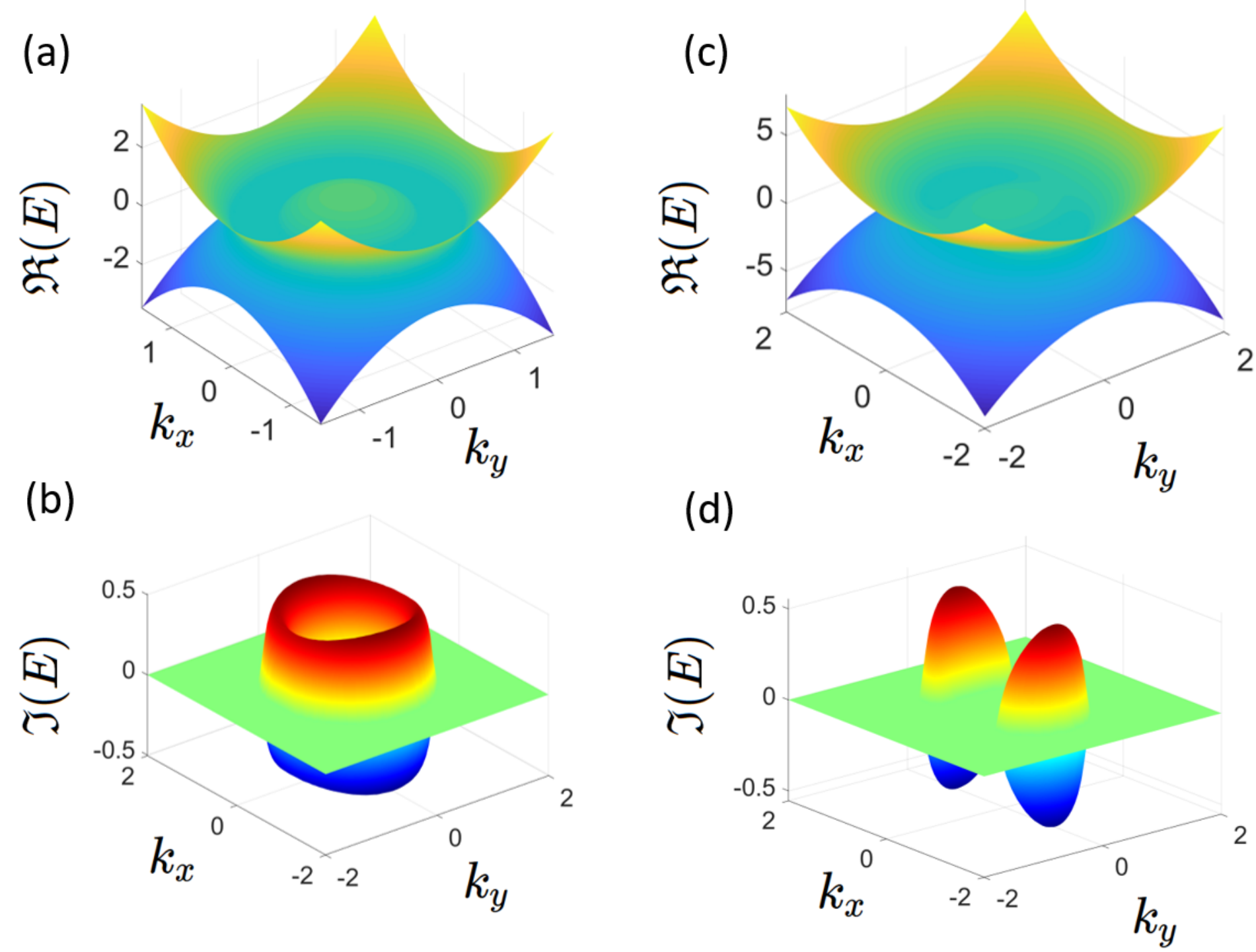}
	\caption{The real part and the imaginary part of the energy dispersion in $k_z=0$ plane for (a,b) $\omega/A=10$ and (c,d) $\omega/A=3$. }
	\label{fig6}
\end{figure}

We illustrate the energy dispersion in the $k_z=0$ plane for the model considered in Fig. \ref{fig6}. In the high-frequency limit, the energy is  purely real both inside the inner ER and outside the outer ER, while it is purely imaginary between the two ERs. As the dipolar pair forms, the energy becomes purely real only outside the two ERs, as shown in Fig. \ref{fig6}(c) and \ref{fig6}(d).

\section{Full Floquet Hamiltonian}\label{app_fullspc}
\begin{figure}[htbp]
	\centering
	\includegraphics[width=\textwidth]{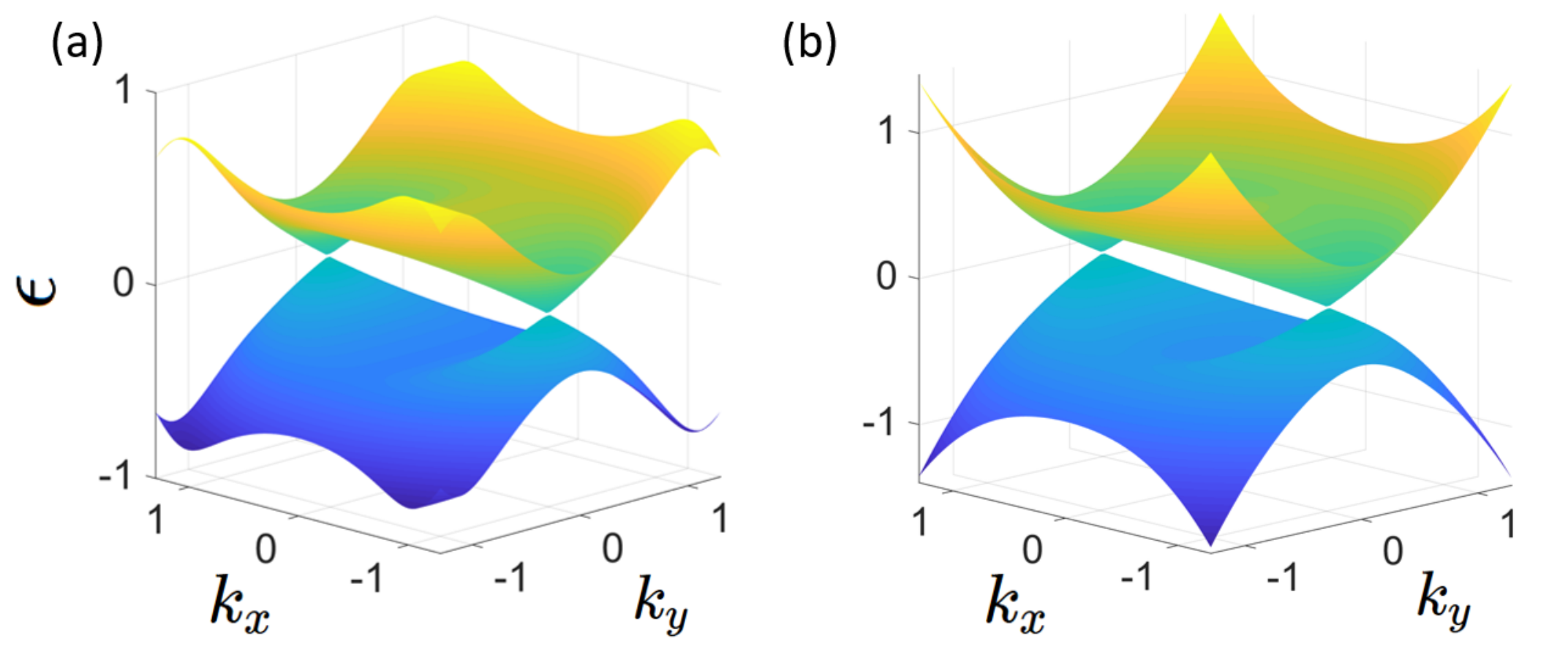}
	\caption{(a) The quasienergy band structure as computed from the full Floquet Hamiltonian (\ref{eq_fullh}) for $k_z=0$. (b) The band  structure for first-order solution from Hamiltonian (\ref{eq_flhm}) for $k_z=0$. The parameters are chosen as $\tilde{m}=1$, $\omega=2$, $\gamma=0$. Energy unit is taken to be 1.}
	\label{fig8}
\end{figure}
One may expect that the first-order Hamiltonian Eq. (\ref{eq_flhm}) might still work well near the Fermi level in the lower-frequency regime due to the diminishing contributions of small high-order momentum components $\mathbf{k}^m$ in the $1/\omega^n$ terms. To examine if the first-order solutions are useful, we compute the quasi-energy spectra from the full Floquet Hamiltonian. The full Floquet Hamiltonian is defined as
\begin{equation}
\hat H_F=\frac{i}{T}\log[\mathcal{T}e^{-i\int_0^T dt \hat H_\mathbf{k}(t)}]\,.\label{eq_fullh}
\end{equation}
For simplicity, we consider only the Hermitian case as an example. The results are shown in Fig. \ref{fig8}. In the Hermitian limit, the band structure computed from the first-order Hamiltonian has qualitative agreement with that of the full Floquet Hamiltonian, where two Weyl points appear in the energy spectra.

\section{Relating the Chern number with the winding number}\label{app_chern}

\begin{figure}[htbp]
	\centering
	\includegraphics[width=\textwidth]{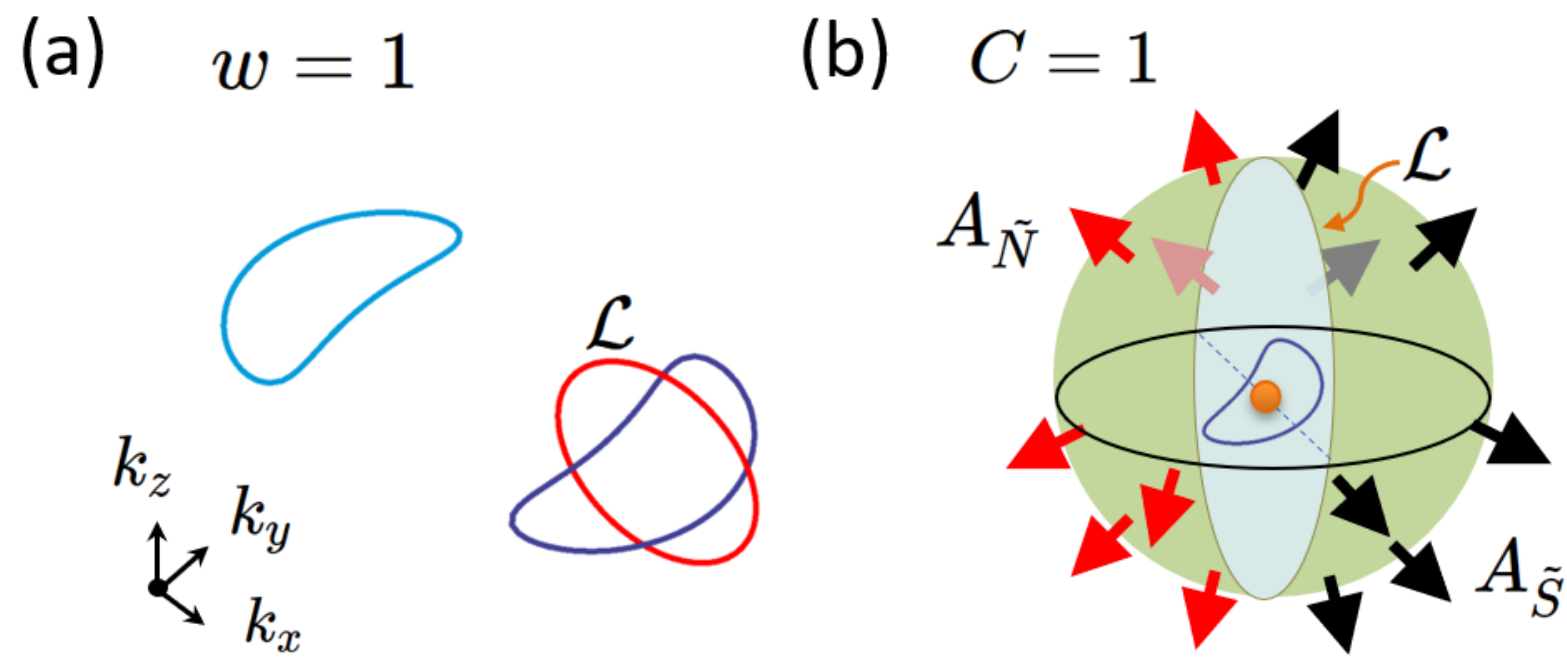}
	\caption{(a) The winding number $w=1$ defined on a loop encircling the ER. (b) The two patches of the gauge are singularity-free.}
	\label{fig8}
\end{figure}

First we perform a basis change $(\tau_y,\tau_z)\mapsto (\tau_z,\tau_y)$, which does not alert the topology. The transformed Hamiltonian is denoted as $\tilde{H}_k=\tilde{\mathbf{d}}\cdot \boldsymbol{\sigma}$, with $\tilde{\mathbf{d}}=(d_x,d_z,d_y)$. Then we map the Hamiltonian $\tilde{H}_k$ to a normalized vector $\vec{n}(\boldsymbol{k})=(\sin (\theta_{i}) \cos (\phi_{j l}), \sin (\theta_{i}) \sin (\phi_{j l}), \cos (\theta_{i}))$. The eigenstates for the lower band can be written as
\begin{equation}
\begin{split}
\langle \tilde{u}_{-}(\theta_{i}, \phi_{j l})|=&[ \cos (\frac{\theta_{i}}{2}), e^{-i \phi_{j l}} \sin (\frac{\theta_{i}}{2})]\,,\\
|u_{-}\left(\theta_{i}, \phi_{j l})\right\rangle=&\left[\begin{array}{c}
 \cos (\frac{\theta_{i}}{2})\\
e^{i \phi_{j l} } \sin (\frac{\theta_{i}}{2})
\end{array}\right]\,,
\end{split}
\end{equation}
which have $\phi$-ambiguity at $\cos (\theta_{i})=\tilde{d}_{i}(\boldsymbol{k}_{0}) /|\tilde{\mathbf{d}}(\boldsymbol{k}_{0})|=\operatorname{sgn}(\Re[h_{i}(\boldsymbol{k}_{0})])=1$. This means that the Berry connection $\mathbf{A}_{\mathbf{k}}=i\langle \tilde{u}|\partial_{\mathbf{k}}|u\rangle$ can not be globally well defined on the integral surface $S^2$. However, one can switch the gauge on the intersection $\mathcal{L}$ of the two hemispheres, as illustrated in Fig. \ref{fig8}(b):
\begin{equation}
\mathbf{A}_{\tilde{S}}=\mathbf{A}_{\tilde{N}}-\nabla\phi_{il}\,.
\end{equation}
Now the Stokes theorem can be applied for an integration over the whole sphere, 
\begin{equation}
\begin{split}
C=&\frac{1}{2 \pi} \oint_{\mathcal{S}} \mathbf{\Omega}_{\theta}(\boldsymbol{k}) \cdot d \boldsymbol{S}\\
=&\oint_{\mathcal{L}} d \ell \cdot (\mathbf{A}_{\tilde{N}}-\mathbf{A}_{\tilde{S}})=w\,.
\end{split}
\end{equation}
This relates the Chern number to the winding number defined in Eq. (\ref{eq_wn}).

\section{Hall conductivity for the Floquet Hamiltonian}\label{Appf}
For the Floquet Hamiltonian $\mathcal{H}(\mathbf{k},t)$ after noise average, the states $|\Psi_\alpha(t)\rangle$ satisfying the Schr$\rm\ddot{o}$dinger equation $i\partial_t |\Psi_\alpha(t)\rangle=\mathcal{H}(\mathbf{k},t)|\Psi_\alpha(t)\rangle$ can be written as $|\Psi_\alpha(t)\rangle=e^{i\epsilon_\alpha t}|\Phi_\alpha(t)\rangle$, due to the time periodicity of the system. We can expand $|\Phi_\alpha(t)\rangle=\sum_n e^{in\omega t}|\Phi_\alpha^{(n)}(t)\rangle$, which satisfies
\begin{equation}
\mathcal{H}_{\mathrm{eff}}(\mathbf{k}) |\Phi_{\alpha}^{(0)}(\mathbf{k})\rangle=E_{\alpha}(\mathbf{k}) |\Phi_{\alpha}^{(0)}(\mathbf{k})\rangle\,,
\end{equation}
\begin{equation}
|\Phi_{\alpha}^{(n)}(\mathbf{k})\rangle=-\frac{\mathcal{H}_{n}}{n \omega} |\Phi_{\alpha}^{(0)}(\mathbf{k})\rangle \quad \text { for } n \neq 0\,.
\end{equation}
The Hall conductivity can be calculated following the linear response theory,
\begin{equation}
\sigma_{yz}=\lim _{\omega \rightarrow 0} \frac{i}{\omega+i 0^{+}}(K_{yz}(\omega)-K_{zy}(0))\,,
\end{equation}
with  $K_{yz}$ the current-current correlation function,
\begin{equation}
\begin{split}
K_{yz}(\omega)=&\sum_{\mathbf{k},m} \int \frac{d \epsilon d \epsilon^{\prime}}{\pi^{2}} \frac{n_{m,\mathrm{F}}\left(\epsilon^{\prime}\right)-n_{m,\mathrm{F}}(\epsilon)}{\epsilon^{\prime}-\epsilon+\omega+i 0^{+}} \times\\
&\operatorname{Tr} (\hat{J}_{y} A_m(\epsilon) \hat{J}_{z} A_m (\epsilon^{\prime} ) )\,,
\end{split}
\end{equation}
where $n_{m,\mathrm{F}}(\epsilon)$ is the distribution function, $A_m(\epsilon)=\Im \rm{Tr}G_m(\omega,\mathbf{k})$ is the spectral function from the retarded Green's function $G_m=\sum_{\alpha=\pm}|\Phi_\alpha^m\rangle\langle \tilde{\Phi}_\alpha^m|/(\epsilon-\epsilon_{\alpha})$, and $\hat J_{y,z}=\frac{\partial \Re\hat H_\mathbf{k}}{\partial k_{y,z}}$ is the current operator.
In this work, we are only interested in the case for the off-resonant driving, since the Lifshitz transition usually happens for large driving frequency. Here we consider only the long-time evolution. The state is close to equilibrium, and only the lower band is occupied. Thus we have a occupation $n_{m,F}=\delta_{m,0}\delta_{\alpha,-}n_{F}(\epsilon)$ with $n_{F}(\epsilon)=1/(e^{\beta\epsilon}+1)$. A similar approximation is used in Ref. \cite{Wang2016}.

\end{appendix}

\end{document}